\documentclass[twocolumn,floatfix,aps, prx]{revtex4-1}

\usepackage{graphicx}
\usepackage{mathtools}
\usepackage{amsmath}
\usepackage{amssymb}
\usepackage{bm}
\usepackage{color}
\usepackage{comment}
\usepackage[FIGTOPCAP]{subfigure}
\usepackage[]{footmisc}
\usepackage{dsfont}
\usepackage{soul,xcolor}
\usepackage[pdfborder={0 0 0}]{hyperref}
\usepackage{lineno}

\newcommand{\vk}{{\bm{k}}}
\newcommand{\bk}{\bm{K}}

\newcommand{\vs}{\boldsymbol{\sigma}}
\newcommand{\vn}{\boldsymbol{\nabla}}
\newcommand{\vB}{\bm{{\cal B}}}

\newcommand{\ZZ}{\mathbb{Z}}

\begin{document}

 
\title{Fermi-arc metals}

\author{Maxim Breitkreiz}
\author{Piet W.\ Brouwer}
\affiliation{Dahlem Center for Complex Quantum Systems and Fachbereich Physik, Freie Universit\" at Berlin, 14195 Berlin, Germany}

\date{March 2023}

\begin{abstract}
We predict a novel metallic state of matter that emerges in a 
Weyl-semimetal superstructure with spatially varying Weyl-node positions.
In the new state, the Weyl nodes  are stretched  into extended, anisotropic Fermi surfaces, which can be understood as being built from Fermi arc-like states.
This ``Fermi-arc metal'' exhibits the chiral anomaly of the parental Weyl semimetal. However, unlike in the parental Weyl semimetal, in the Fermi-arc metal the ``ultra-quantum state'', in which the anomalous chiral Landau level is the only state at the Fermi energy, is already reached for a finite energy window at zero magnetic field. The dominance of the ultra-quantum state implies a universal low-field  ballistic magnetoconductance and the absence of quantum oscillations, making the Fermi surface ``invisible'' to de Haas-van Alphen and Shubnikov-de Haas effects, although it signifies its presence in other response properties.

\end{abstract}
\maketitle

\emph{Introduction}---The characteristic property of 
Weyl semimetals (WSMs) \cite{Wan2011,Burkov2011,Xu2011,Xu2015, Xu2015b, Lv2015,Armitage2017, Yan2017, Burkov2017,Hasan2021, Bernevig2022a}  is the presence of Weyl nodes, topologically protected band-degeneracy points with a diabola-shaped dispersion, near the Fermi energy.
The Weyl nodes may be assigned a chirality $\chi = \pm 1$, which manifests itself in the chiral anomaly: In an applied magnetic field, the Weyl nodes turn into Landau levels, with an anomalous chiral Landau level propagating parallel ($\chi = 1$) or antiparallel ($\chi = -1$) to the applied magnetic field \cite{Adler1969,Bell1969}. 
In WSMs, Weyl nodes occur in pairs of opposite chirality, which, on the crystal surface, 
reconnect in momentum space via Fermi-arc surface states  \cite{Armitage2017}.
The chiral Landau levels and Fermi arcs are responsible for a range of characteristic transport effects of WSMs, such as an anomalous Hall effect and a negative magnetoresistance in parallel electric and magnetic fields \cite{Burkov2017, Breitkreiz2019, Piskunow2021}.

Building on the increased availability of high-quality WSM materials,
there is currently a growing interest in the manipulation of Weyl-node positions in  energy-momentum space \cite{Bouhon2020, Chen2021a}. Particularly interesting is the proposed and observed dependence of the Weyl-node positions in magnetic WSMs on the magnetization direction \cite{Cano2017,Nie2020,Cheng2022,Sarkar2022}.
This opens the possibility of WSM superstructures with spatially modulated node positions, either based on heterostructures, or based on intrinsic node-position variations from a magnetic texture. Helical magnetic order can be favored by Weyl-Fermion-mediated Ruderman-Kittel-Kasuya-Yosida interactions between local magnetic moments \cite{Chang2015a,Hosseini2015,Wang2017a, Park2018, Nikolic2021} and has been recently demonstrated experimentally in NdAlSi \cite{Gaudet2021} and SmAlSi \cite{Yao2022}. An ultrathin variant of a helical magnetic texture is also found in Bloch-type domain walls of Co$_3$Sn$_2$S$_2$ \cite{Lee2021}, where the magnetization reverses its direction via a helical half turn.
Theoretically, such WSM superstructures have remained widely unexplored.

\begin{figure}[b]
\includegraphics[width=\columnwidth]{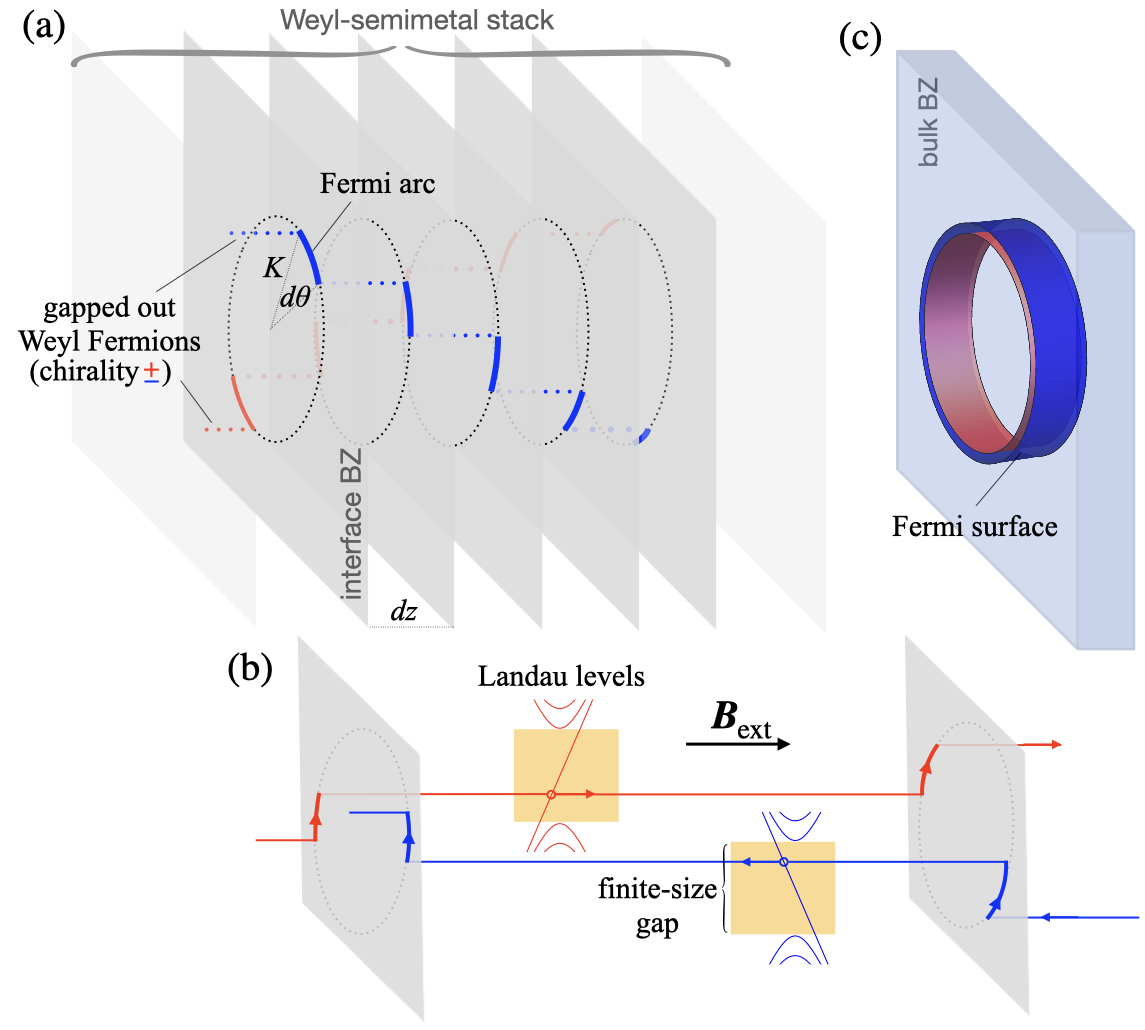}
\caption{(a) Fermi-arc metal from a stack of 
WSMs with  Weyl-node positions (dotted red/blue lines for Weyl-node of chirality $+/-$) rotating discretely in form of a double helix, shown in a mixed real
($z$) and momentum ($k_x,k_y$) space. Interfaces host Fermi arcs (red/blue lines) that connect Weyl nodes of the same chirality at opposite sides of the interface.
(b) An applied magnetic field $\bm{B}_{\rm ext}$ along the stacking direction leading to particle flow along Fermi arcs on the interface and via anomalous chiral Landau levels of Weyl Fermions between interfaces. The propagation direction (parallel or antiparallel to $\bm{B}_{\rm ext}$) depends on the chirality of the Weyl node. Higher Landau levels are finite-size-gapped due to reflections at interfaces.
(c) Reduced bulk Brillouin zone with cylindrical chiral Fermi surfaces (red/blue for chirality $+/-$). }
\label{figi}
\end{figure}

In this work we show that WSMs with a periodic spatial modulation of the Weyl node positions are in a novel semimetallic state, which we propose to call  a ``Fermi-arc metal'', since its low-energy spectrum can be interpreted as being entirely built from Fermi arcs, as we explain below. The Fermi-arc metals constitute a realization of the chiral anomaly, but with essential differences compared to the WSMs: The Weyl nodes are replaced by two cylindrical Fermi surfaces, defined in the reduced Brillouin zone corresponding to the spatial modulation of the Weyl node position.  The Fermi surfaces have a well-defined chirality and they give rise to characteristic  
anomalous response properties.

\emph{Heterostructure model}---Before giving a generic formulation  in the subsequent sections, we will first provide basic insight into the Fermi-arc metal 
 from a  specific construction consisting of a stack of WSMs, illustrated in Fig.\ \ref{figi}(a). 
To keep the discussion simple, each layer is taken to consist of a minimal WSM with a single pair of Weyl nodes at momenta $\pm \bm{K}$, so that for momenta $\vk$ in the vicinity of the Weyl nodes, the low-energy Hamiltonian is
\begin{equation}
  H = \pm \vs \cdot (\bm{k} \mp \bm{K}).
\end{equation}
(Here $\pm$ denotes the chirality and $\vs$ is the vector of Pauli matrices.) We use the coordinate $z$ for the stacking direction and take the width of each layer to be $dz$. We assume that the component $\pm \bm{K}_{\perp}$ of the node positions perpendicular to the stacking ($z$) direction is nonzero and that the node positions are rotated by an angle $d\theta$ between adjacent layers, such that they form a double helix with radius $K_{\perp}$ and the period length (pitch) $2\pi/Q = 2\pi dz/d\theta$, see Fig.\ \ref{figi}(a). We will further assume that $Q \ll K_{\perp}$, {\em i.e.}, the pitch is much larger than the inverse momentum-space separation of the Weyl nodes.

We'll now argue that, at low energies, bulk states in this WSM stack are built from Fermi-arc states residing at the interfaces between layers, whereas states associated with the bulk Weyl nodes in the layers are gapped out. We first consider the Fermi-arc interface states. Hereto, we note that an interface between two WSMs with displaced 
Weyl node positions hosts Fermi arcs that connect projections of Weyl nodes of the \emph{same} chirality on the two opposing sides of the interface
\cite{Dwivedi2018,Abdulla2021,Mathur2022,Kaushik2022,Buccheri2022,Chaou2023}, 
 which follows from the requirement that the Fermi arcs should vanish 
 in the limit $d\theta \to 0$ of a homogeneous 
Weyl semimetal. 
For generic momenta, Fermi arcs are exponentially localized at the interface. For momenta near the (projected) Weyl node of one of the adjacent layers, the Fermi arc states extend into that layer, but not farther, because of the shifted Weyl node position in the next layer.
The smallest (real-space) distance between two Fermi arcs of different chirality at the same momentum is of the order of  half the pitch $\pi/Q$. Hence, since $Q \ll K_{\perp}$, the heterostructure preserves chirality up to the exponentially small
(in $K_{\perp}/Q$) overlap of interface arcs. 
Now turning to the bulk Weyl nodes  we note that these, too, have 
a finite extension $\Delta z$, because the Fermi pockets do not overlap for layers that are sufficiently far apart. At energy $E$ one has $\Delta z \sim E/Q K_{\perp}$, so that the finite-size gap $\sim 1/\Delta z$ removes the bulk Weyl Fermions at energies $E^2 \lesssim QK_{\perp}$.  We thus find that for $E^2 \lesssim Q K_{\perp}$ the stack of WSMs with modulated Weyl node positions has a Fermi surface that is built entirely from the chiral Fermi arcs. This is the ``Fermi-arc metal''. 
In the presence of a magnetic field perpendicular to the interfaces, the gapping only applies to higher Landau levels, while the anomalous lowest Landau level remains to reconnect the Fermi arcs, as illustrated in Fig.\ \ref{figi}(b).

This simple model of a Fermi-arc metal allows us to identify two key features of its Fermi surface: The Fermi surface consists of two approximately cylindrical sheets, which are hole- and electron-like depending on the chirality, as illustrated in Fig.\ \ref{figi}(c). The approximately cylindrical shape follows from the localization to interfaces in the stacking direction and the rotational symmetry around it. The electron- and holelike character of the Fermi surfaces can be identified by considering the particle flow through the heterostructure in the presence of a magnetic field in the stacking direction, illustrated in Fig.\ \ref{figi}(b).  In a more general setting, to be discussed below, the rotation symmetry may be lifted, but the other two properties (flat dispersion in the modulation direction, well-defined chirality for each Fermi surface) remain.

From the topological point of view, the persistence of Fermi arcs despite the removal of the Weyl nodes is possible and even necessary because according to the WSM topology, band touchings can only disappear pairwise, via coupling of opposite chiralities. In standard WSMs, the breaking of translation invariance couples chiralities, so that Weyl Fermions and their chirality disappear simultaneously, as it happens at the WSM surface. The layer construction breaks translation invariance (on the scale of the unit cell of the individual layers) in a chirality-preserving way, allowing to eliminate Weyl nodes while preserving the chirality, which is transferred to the Fermi arcs. The persistence of 
chirality is signified by the persistence of the chiral anomaly (see Fig.\ \ref{figi}(b) and, for a general setting, below).

\emph{Weyl-node stretching}---We now consider a generic chirality-preserving deformation of a Weyl node into a Fermi arc-like state. We first consider the Hamiltonian of a single Weyl Fermion with positive chirality,
\begin{align}
H = \sigma_z k_z+ \vs_{\perp} \cdot\big[ \vk_{\perp} -  \bk_{\perp} (z)\big],
\label{h0}
\end{align}
where the node-position $\bk_\perp(z)$ varies along the $z$ direction and we abbreviated $\vs_{\perp} = (\sigma_x,\sigma_y)$, $\vk_{\perp} = (k_x,k_y)$. To keep the discussion simple, we do not consider variations of the longitudinal node position $K_z$, which would only modify the phase of the wavefunction, a possible velocity anisotropy of the Weyl node (here velocity is set to one), which would effectively rescale the momenta, and any momentum-dependent (pseudo-) scalar potentials, which would tilt the Weyl cone but otherwise do not modify the results in a qualitative manner.

For a smooth node-position variation (precise condition to be specified below) low-energy states at a transverse momentum $\vk_{\perp}$ will be near those $z$ for which $\vk_{\perp}$ is close to the node position $\bk_{\perp}(z)$. 
  Let $z_{\vk_{\perp},n}$, $n \in \ZZ$ denote all local minima of 
$|\bk_{\perp}(z)-\vk_{\perp}|$. Since $\bk_{\perp}(z)$ enters the Hamiltonian (\ref{h0}) as a vector field, expanding to first order in $z$ around a minimum $z_{\vk_{\perp},n}$, the Hamiltonian becomes that of a Weyl Fermion in an effective (pseudo) transverse magnetic field 
$\vB (z_{\vk_{\perp},n}) = \vn\times \bk_{\perp}(z_{\vk_{\perp},n})$ 
 \cite{Ilan2020}. The energy levels of a Weyl Fermion in such a magnetic field are well known: The lowest energy level $E_{\vk_{\perp},n}$ is the corresponding anomalous Landau level, which is linear in the momentum component along the field direction, 
\begin{equation}
E_{\vk_{\perp},n}=   [ \vk_{\perp} -  \bk_{\perp} (z_{\vk_{\perp},n})]\cdot \hat {\vB} (z_{\vk_{\perp},n}), 
\label{e0}
\end{equation}
where $\hat{\vB} (z)=\vB (z)/|\vB(z)|$ is the unit vector pointing in the direction of $\vB(z)$. The wavefunction reads
\begin{equation}
  \psi_{\vk_{\perp},n}(z) =
  \eta_{\vk_{\perp},n} \,
  e^{-|\vB(z_{\vk_{\perp},n})|
      (z-z_{\vk_{\perp},n})^2/2},\label{wv}
\end{equation}
with $\eta_{\vk_{\perp},n}$ the eigenvalue-one eigenspinor of $\vs\cdot \hat{\vB} (z_{\vk_{\perp},n})$.
The eigenfunctions defined this way form an orthonormal basis if the wavefunctions at different $n$ don't overlap, {\em i.e.}, if $\min_{m\neq n}|\vB (z_{\vk_{\perp},n})|(z_{\vk_{\perp},m}-z_{\vk_{\perp},n})^2 \gg 1$.
Consistent with the localization of wavefunctions, the dispersion is flat ($k_z$-independent) in the longitudinal direction. 
The gap to higher energy levels is $\sqrt{2|\vB(z_{\vk_{\perp},n})|}$.
What turns the WSM into a Fermi-arc metal can thus be interpreted as an effective momentum-dependent magnetic field $\vB(z_{\vk_{\perp},n})$, which stretches the Weyl node into the chiral Landau level, the latter being the known equivalent of the Fermi arc \cite{Ilan2020}.

For the specific example of a helical node-position variation, $\bk_{\perp}(z)=K_{\perp}(\cos Qz,\, \sin Qz)$, the requirements for a Fermi-arc metal are met for $Q\ll K_\perp$ and we obtain $E_{\vk_{\perp},n}=-(|\vk_{\perp}|-K_{\perp})$ and $\vB (z)=-Q \bk_{\perp}(z)$.
This gives a cylindrical, holelike Fermi surface, separated from higher bands by $\sqrt{2QK_\perp}$.

The full Hamiltonian must also contain the Weyl node of the opposite (negative) chirality, as well as possible other states. The spatial localization in the $z$ direction allows a spatial separation of different degrees of freedom. In the minimal model of a helical magnetic order in the magnetic WSM with two Weyl nodes discussed above, the opposite-chirality node is naturally placed at the momentum $-\bk_{\perp}(z) = \bk_{\perp}(z+\pi/Q)$, so that the two chiral nodes form a double helix. The wavefunction centers then are at
\begin{equation}
z_{\vk_{\perp},n,\chi} = \frac{\theta_{\vk_{\perp}} + \pi \delta_{\chi,-}+2\pi n}{Q},\label{zi}
\end{equation}
where $\chi=\pm 1$ is the chirality and $\theta_{\vk_{\perp}}$ the angle indicating the direction of $\vk_{\perp}$. Because of the spatial separation, the coupling between chiralities is exponentially small in $K_{\perp}/Q$. Since the anomalous Landau level of the opposite chirality moves in the opposite direction, the full dispersion is 
\begin{equation}
E_{\vk_{\perp},n,\chi} = -\chi (|\vk_{\perp}|-K_{\perp}),
\end{equation}
which gives an additional electronlike cylindrical Fermi surface for $\chi=-1$. There is a cylindrical surface of degeneracy at zero energy, protected (up to corrections exponentially small in $Q K_{\perp}$) by the spatial separation of states of opposite chirality.
The results for the continuous helical node-separation variation are in full quantitative agreement with what we found for the discrete heterostructure model in the limit $d \theta \ll 1$.

\begin{figure}
\includegraphics[width=0.9\columnwidth]{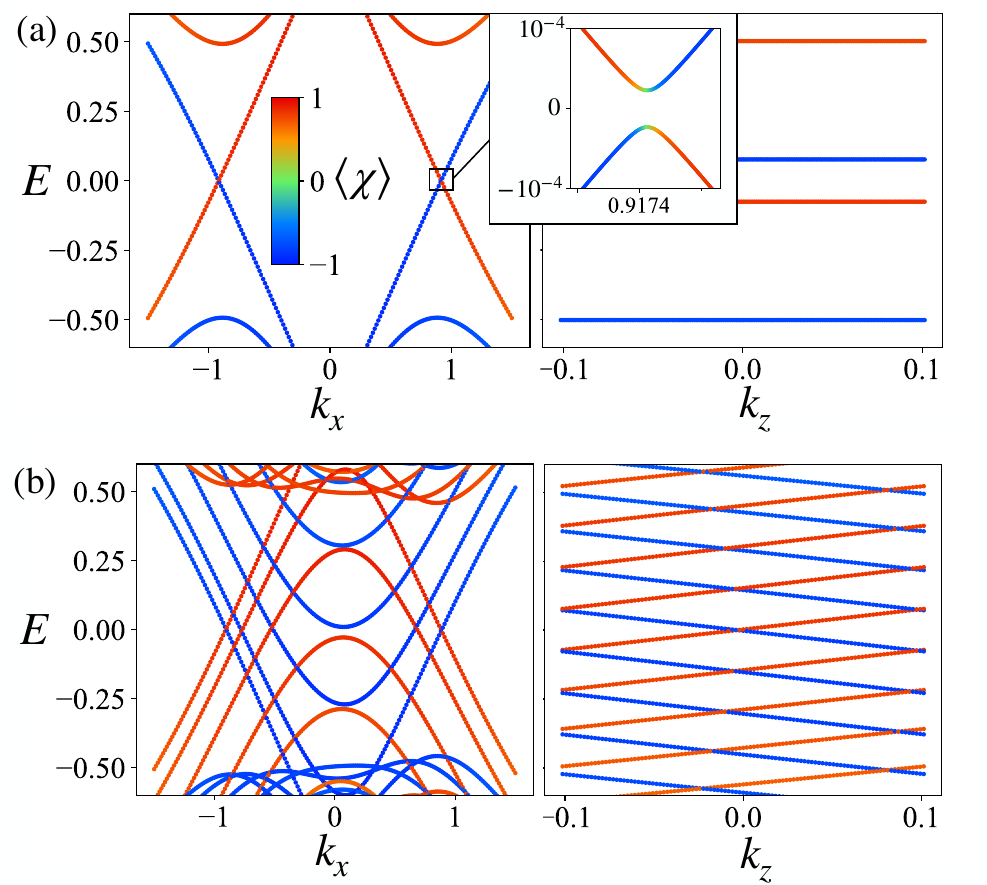}
\caption{Dispersion from numerical 
diagonalization of a lattice model, which at low energy 
corresponds to the continuum model with nodes at
$\pm \bk_\perp(z) =\pm( \cos(Qz)\hat{y} -\sin (Qz)\hat{x})$,  
 $Q=2\pi/31$ \cite{dummy2}
(a) as a function of $k_x$ at $k_y=0=k_z$ (left) and as a function of $k_z$ at $k_y=0,\, k_x=1$. Color encodes the expectation value of the chirality. The inset shows a close-up at the crossing at zero energy, where chirality conservation is violated for energies exponentially small in $K_{\perp}/Q$. The dispersion is rotation-symmetric around
$k_z$. 
(b) 
Same as (a) but with an applied magnetic field $\bm{B}_\mathrm{ext}=\hat{x} \, 2\pi/(10\times 31)$ (left) and $\bm{B}_\mathrm{ext}=\hat{z} \, 2\pi/30$ (right). (Normalization of the applied field is chosen such, that $|\bm{B}_\mathrm{ext}|$ is the inverse square of the magnetic length in units of the lattice constant.)  Note that the right panel shows a single 
Landau level per chirality, backfolded into the reduced Brillouin zone ($k_z\in [-Q/2,Q/2]$).}
\label{fignum}
\end{figure}

Figure \ref{fignum}(a) shows the results of a numerical calculation of the dispersion of a Fermi-arc metal based on a tight-binding model of a magnetically doped topological insulator \cite{Qi2011,Fu2010a, Liu2013b, Cho2011, PhysRevLett.110.126804}, in the presence of a helical magnetization rotation 
(for details see Supplemental Material (SM) 
\cite{*[{See Supplemental Material for details of the 
lattice simulation of the Fermi-arc metal}] [{}] dummy2}). 
The numerics confirms the analytically calculated dispersion and the exponentially suppressed coupling of chiralities at the degeneracy points. 

\emph{Chiral anomaly and surface states}---The elimination of Weyl nodes, while preserving chirality of Weyl Fermions, opens the question, whether and how the Fermi-arc metal shows the chiral anomaly, {\em i.e.}, the non-conservation of chiral charge.
In the heterostructure model, the persistence of the chiral anomaly follows from the persistence of chiral Landau levels between the interfaces, cf.\ Fig.\ \ref{figi}(b). More generally, it can be understood by considering a WSM with two Weyl nodes, of which only one has a $z$-dependent node position. In this case, the momentum-space position of the other Weyl node is constant, so that this node hosts a conventional Weyl Fermion, which is subject to the standard chiral anomaly. 
Then the non-conservation of chiral charge in the Fermi-arc metal corresponding to the node with $z$-dependent momentum-space position follows from the conservation of total charge and the non-conservation of charge by the conventional Weyl Fermion.  

The lattice simulation confirms the presence of the chiral anomaly in the Fermi-arc metal. Figure \ref{fignum}(b) shows that for any field direction, parallel and perpendicular to $z$, each chirality has one unbalanced Landau level, which moves parallel or antiparallel to  the field direction, depending on the chirality.  

The presence of the chiral anomaly implies the presence of Fermi-arc surface states, because these surface states must mediate particle transport between chiralities in a magnetic field normal to the surface.  A numerical simulation is given in the SM 
\cite{dummy2}.

\emph{Ultra-quantum state}---While the non-conservation of chiral charge is thus inherited from the parental WSM, there is a sense in which the manifestation of the chiral anomaly is even stronger in a Fermi-arc metal than in a standard WSM: Whereas in standard Weyl semimetals the so-called ``ultra-quantum state''
\cite{Burkov2015c,lu2022}, in which the anomalous chiral Landau levels are the {\em only} levels at the Fermi energy $E_{\rm F}$, requires a threshold value $ B_{\rm ext} > E_{\rm F}^2/2$ for the applied magnetic field, in a Fermi-arc metal there is a range of Fermi energies $|E_{\rm F}| < \sqrt{2 |\vB|}$ for which the ultra-quantum state already appears at zero magnetic field. This is illustrated in the left panel of Fig.\ \ref{fignum}(a) and the right panel of Fig.\ \ref{fignum}(b). 
(Note that the dispersion at  \emph{low} fields in the $x$ 
(or similarly $y$) direction is equivalent to 
that in the left panel of Fig.\ \ref{fignum}(a) due to the vanishing $k_z$ dispersion, 
 see SM \cite{dummy2}  for details.)
Within the analytical approach, the effect of an applied field $\bm{B}_{\rm ext} = B_{\rm ext} \hat{z}$ can be easily understood by adding $\bm{B}_{\rm ext}$ to the effective field $\chi \vn\times \bk_{\perp}$ used for the derivation of \eqref{e0}. With $\vB = \chi \vn\times \bk_{\perp} +e\bm{B}_{\rm ext}$,  the derivative of the dispersion \eqref{e0} with respect to $k_z$ gives the velocity along the field, $v_z = \chi  B_{\rm ext} / K_{\perp}Q$, which is equal for all states and opposite for opposite chiralities
\footnote{The application of a transverse magnetic field suppresses the energy range of the ultra-quantum state, because such an applied field may partially compensate the effective field $\chi \vn\times \bk_{\perp}$. This can be seen from the numerics by comparing the left panels of Fig.\ \ref{fignum}(a) and (b).}.
 
As a characteristic signature of the extended ultra-quantum state, the longitudinal magneto-conductance 
assumes a universal linear dependence on $B_{\rm ext}$, $G=(e^2/h) N_B$, where $N_B$ is the number of flux quanta through the system, which is also the degeneracy of the anomalous chiral Landau level. This linear dependence holds down to $B_{\rm ext} = 0$, because the chiral Landau level is the only level at the Fermi energy. 
In contrast, in a conventional WSM, the transport is dominated by the zeroth Landau level only above a non-universal threshold field, which depends on the diffusion constant and the density of states \cite{Altland2016, Burkov2017}. 
The maximal system length for this effect to occur is set by the scattering 
length for scattering across the 
Weyl-node separation  \cite{Altland2016}, which in Fermi-arc metal 
 can be expected to be extended, due to  the 
spatial separation of wavefunctions (see disorder effects below).

\emph{Quantum oscillations}---A Fermi-arc metal has a well-defined Fermi surface,
which is closed in the plane perpendicular to the node-variation direction, as 
discussed above. In all conventional metals (including Weyl semimetals) 
such Fermi surfaces  are detectable via quantum-oscillation experiments, such as the de Haas-van Alphen and Shubnikov-de Haas effects \cite{Zhang2005,He2014,Huang2015a,Hu2016,VanDelft2018}.
The oscillations stem from Landau levels passing through the Fermi energy and 
the  semiclassical quantization rule \cite{Roth1966, Gao2017, Breitkreiz2018}, based on 
selfinterference assumption of a wavefunction on a closed orbit, 
relates the oscillation frequency with the Fermi-surface shape  \cite{Ashcroft}.

For the Fermi surface of a Fermi-arc metal one would expect oscillations for $\bm{B}_{\rm ext} = B_{\rm ext} \hat{z}$, as the momentum-space intersection of the Fermi surface and a plane at constant $k_z$ is a closed orbit 
(a circle with area $\pi K_{\perp}^2$ for the helical case).
Yet, no Landau levels pass through the Fermi energy if $B_{\rm ext}$ is varied, because the Fermi-arc metal is in the ultra-quantum state, as discussed above.
The failure of the semiclassical quantization rule for Fermi-arc metals can also be understood semiclassically from the anomalous motion of the wavefunction: Taking again the helical case as an example, the Lorentz force, $d\vk_{\perp} = -\chi \hat k_{\perp} \times \bm{B}_\mathrm{ext} \ dt$, moves a particle around the cylinder at a fixed $k_z$ in the time  $\Delta t=  2\pi\, K_{\perp} /B_{\rm ext}$. (Here $\hat k_{\perp} = \vk_{\perp}/|\vk_{\perp}|$ is the unit vector in the direction of $\vk_{\perp}$.) During this time, the wavefunction localization center moves the distance $\Delta z= \chi 2\pi/Q$ along the applied field, see \eqref{zi}. (The same result follows from the anomalous velocity $v_z = \chi B_{\rm ext}/ K_{\perp} Q$ derived above based on the full quantum-mechanical model). 
This anomalous shift prohibits the self-interference and thus invalidates the application of the semiclassical quantization rule. 

In a slab of width $L$  in the $z$ direction, the wavefunction motion becomes closed by the Fermi-arc surface states into an orbit going through the whole slab, similar to
 the Weyl orbit of a conventional WSM \cite{Potter2014, Moll2016}. However, unlike 
the Weyl orbit, each $z$-shift  of $2\pi/Q$ is associated 
with an enclosed momentum-space
 area of the cylindrical Fermi surface, as discussed above, and the oscillation frequency 
 is thus proportional to $L$. 
In samples with non-parallel surfaces or a large width, $L\gg Q$, the oscillations
 stemming from this orbit will average out or will be strongly suppressed by decoherence effects,  owing to the large orbit length.

 The absence of quantum oscillations (in bulk materials) becomes a striking signature of a clean Fermi-arc metal in connection with other measurements signifying the presence of the Fermi surface and the purity of the sample, such as angle-resolved photo-emission spectroscopy and various transport measurements. We are not aware of the existence of another metal with this property.

 \emph{Disorder and interaction effects}---The spatial localization of wavefunctions in Fermi-arc metals poses a number of intriguing questions with regard to the effects of disorder and electron-electron interactions, a complete discussion of which goes beyond the scope of this work.
In diffusive transport regimes one can expect an enhancement of the conductivity along the dispersive (transverse) directions, as the spatial separation of counterpropagating states prohibits disorder-induced backscattering. In particular, the  scattering angle for a helical node-position variation is limited by the wavefunction overlap to $\Delta \theta\sim \sqrt{Q/K_{\perp}}$ and the relaxation must go via multiple scattering events, even when the disorder potential is short-ranged.
On the other hand, the vanishing spatial overlap of wavefunctions does not hinder Coulomb interaction. In particular, the possibility of an overlap of chiralities in the energy-momentum space can favor interaction-driven spontaneous breaking of chiral symmetry.
This is known to lead to the sought-after condensed-matter realization of axions \cite{Wang2013, Gooth2019, Shi2021}, which, among other things, could play a crucial role in the detection of dark matter \cite{Preskill:1982cy,Abbott:1982af,Dine:1982ah,Marsh2019,Chigusa2021}.
The appearance of Weyl-Fermion chirality in the new form of a Fermi-arc metal opens new routes in this direction.

\emph{Acknowledgments}---We thank Max Geier, Carsten Timm, Tommy Li, and Christiane P.\ Koch for fruitful discussions.
This research was supported by project A02 of the CRC-TR 183 “entangled states of matter” and Grant No.\ 18688556 of the Deutsche Forschungsgemeinschaft (DFG, German Science Foundation).

\bibliography{library}

\renewcommand{\theequation}{S\arabic{equation}}
\renewcommand{\thefigure}{S\arabic{figure}}

\setcounter{equation}{0}
\setcounter{figure}{0}



\onecolumngrid

\vspace*{10cm} 

\section*{Supplemental Material}

\subsection*{Lattice simulation of the Fermi-arc metal}

\twocolumngrid

\begin{figure}[b]
\includegraphics[width=\columnwidth]{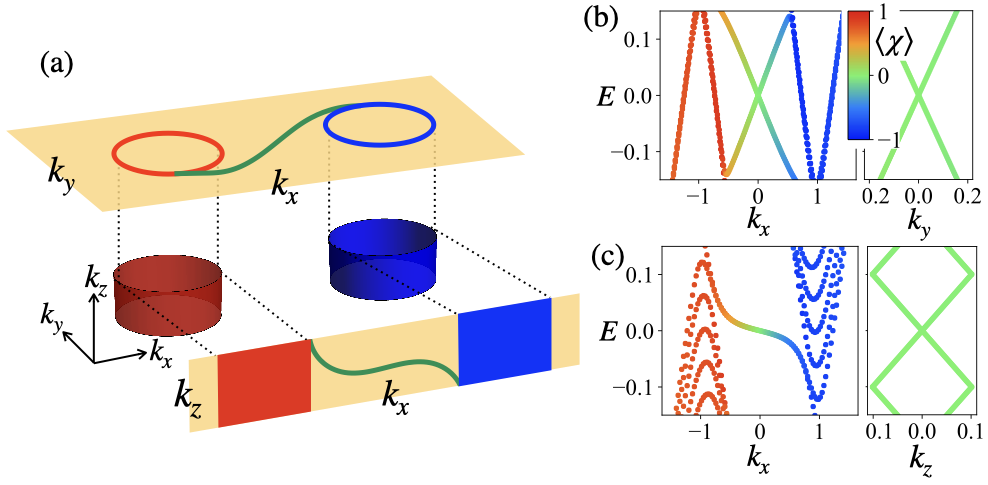}
\caption{Topological surface states of a Fermi-arc metal. (a) Schematic illustration of Fermi arcs (green) connecting projections of the bulk cylindrical Fermi surfaces of the two chiralities (blue/red) in the surface Brillouin zones of the $xy$ surface and the $xz$ surface. (b) Dispersion
[$k_x$-dependence at $k_y=0$ (left) and 
$k_y$-dependence at $k_x=0$ (right)] from numerical diagonalization of the lattice model \eqref{latmodel} with $\bk_\perp(z) = (1+0.3\cos (Qz))\hat{x}+0.3\sin(Qz)\hat{y}$,  $m=0.1,\, Q=2\pi/31\,$, and $t=1=t'$ for a slab with a finite $z$-width $L_z=n\times 31,\, n=3$ (no visible difference in changing $n$ since the states are flat in $k_z$). (c) Same as (b) but for a finite $y$-width $L_y=40$.}
\label{figss}
\end{figure}

For the lattice simulation shown in Figs.\ \ref{fignum} and \ref{figss} we consider a Fermi-arc metal built from the paradigmatic model of a magnetically doped three-dimensional topological insulator \cite{Qi2011,Fu2010a, Liu2013b, Cho2011, PhysRevLett.110.126804}. Specifically, we use a four-band tight-binding model on a cubic lattice, where we include a $z$-dependent ``magnetization'',
\begin{align}
H_\mathrm{tb}  =&\ \sum_{i=x,y,z} [t\, \chi_z\sigma_i \sin k_i 
+t' \chi_x(1-\cos k_i)] \nonumber\\
&\ \ \ \ \ +\vs_{\perp} \cdot\bk_{\perp} (z)+\chi_x m.
\label{latmodel}
\end{align}
Here $\chi_{i}$ and $\sigma_i$ are Pauli matrices, the momentum $k_i$ is in units of the inverse lattice constant, which is set to one, and the size of the unit cell in the $z$ direction, $2\pi/Q$, is taken to be an integer multiple of the lattice constant. 
An external magnetic field $\bm{B}_{\rm ext} =\vn\times\bm{A}_{\rm ext}$ is incorporated via the Peierls substitution, $k_i\to k_i-A_{\mathrm{ext}, i}$, and the introduction of a magnetic unit cell, with a size that is an integer multiple of the size of the unit cell in a direction perpendicular to $\bm{B}_{\rm ext}$. 

The results of numerical diagonalization, confirming the expected dispersion, the exponentially suppressed coupling of chiralities at the degeneracy points, and the persistence of the chiral anomaly, are shown in Fig.\  \ref{fignum}. The parameters 
in the plot are $\bk(z) = \cos(Qz)\hat{y} -\sin (Qz)\hat{x}$,  
 $Q=2\pi/31$, $m=0.1$, and $t=1=t'$.
The dispersion of systems in two different slab geometries, showing topological surface states on surfaces parallel to the $xy$ and $xz$ planes, is shown in Fig.\ \ref{figss}.  Whereas Fermi-arc surface states are guaranteed to exist on a surface parallel to the $xy$ plane because of the spatial separation of the chiralities, uniform protected Fermi-arc surface states on surfaces parallel to the $z$ axis exist only if the Fermi surfaces for opposite chiralities do not overlap.  For this reason, for the calculations in Fig.\ \ref{figss} the $z$-dependence of the node positions $\bk_{\perp}(z)$ was chosen such, that Fermi surfaces of the two opposite chiralities do not overlap in reciprocal space

As the main purpose of the plots in Fig.\  \ref{fignum}(b) is to confirm the chiral anomaly, a large value was chosen for the strength of the applied magnetic field, in order to better distinguish the levels. In particular, the magnetic length  $l_B=1/\sqrt{B}$ was taken to be smaller than the unit cell in the $z$ direction,  $2\pi/Q$. While there is no qualitative difference to the low-field dispersion with regard to the chiral anomaly, 
there is a qualitative difference between the dispersions for $l_B\ll 2\pi/Q$ (strong field, shown) and $l_B\gg 2\pi/Q$ (weak field, not shown) for a magnetic field in the $x$ direction. 
In the latter case, due to the vanishing $k_z$ dispersion on length scales larger than the unit cell $2\pi/Q$, the gap between the Landau levels vanishes, which is in line with the fact that the Fermi surface is open in the $z$ direction.

Also note that similar to the standard WSMs, there is an upper bound for magnetic fields above which the field starts to couple chiralities. This upper bound is set by the minimal transverse distance of nodes in momentum space, which for decoupled chiralities must be larger than the inverse magnetic length. For the helical node-position  variation this means $B_{\rm ext} \lesssim K_{\perp}^2$, which for well-separated Weyl nodes lies at very large fields in realistic materials.

\end{document}